\documentclass[%
 reprint,
superscriptaddress,
unsortedaddress,
 amsmath,amssymb,
 aps, pre,
]{revtex4-2}

\usepackage{graphicx}
\usepackage{dcolumn}
\usepackage{bm}
\usepackage{hyperref}

  \newcommand{\av}[1]{\left\langle#1\right\rangle}
  \newcommand{\cbr}[1]{\left(#1\right)}
  \newcommand{\sbr}[1]{\left[#1\right]}

  \newcommand{\moby}{\mu}
  \newcommand{\frict}{\gamma}
  \newcommand{\mean}{\lambda}
  \newcommand{\tmean}{\widetilde{\lambda}}
  \newcommand{\stiff}{k}
  \newcommand{\tstiff}{\widetilde{k}}
  \newcommand{\stud}{k}
  \newcommand{\tstud}{\widetilde{k}}
  
  \newcommand{\dimless}[1]{{#1}^*}
  
  \newcommand{\levy}{L\'{e}vy }

\DeclareRobustCommand{\vect}[1]{
  \ifcat#1\relax
    \boldsymbol{#1}
  \else
    \mathbf{#1}
  \fi}

\begin{document}

\preprint{}

\title{Shortcuts To Adiabaticity for \levy processes in harmonic traps}

\author{Marco Baldovin}
\affiliation{Universit\'{e}  Paris-Saclay,  CNRS,  LPTMS,  91405,  Orsay,  France}%
 \email{marco.baldovin@universite-paris-saclay.fr}
\author{David Gu\'{e}ry-Odelin}%
\affiliation{Laboratoire  Collisions,  Agr\'{e}egats,  R\'{e}eactivit\'{e}e,  IRSAMC,  Universit\'{e}  de  Toulouse,  CNRS,  UPS,  France}%
\author{Emmanuel Trizac}%
\affiliation{Universit\'{e}  Paris-Saclay,  CNRS,  LPTMS,  91405,  Orsay,  France}%

\date{\today}

\begin{abstract}
\levy stochastic processes, with noise distributed according to a \levy stable distribution, are ubiquitous in science. Focusing on the case of a particle trapped in an external harmonic potential, we address the problem of finding ``shortcuts to adiabaticity'': after the system is prepared in a given initial stationary state, we search for time-dependent protocols for the driving external potential, such that a given final state is reached in a given, finite time.
These techniques, usually used for stochastic processes with additive Gaussian noise, are typically based on a inverse-engineering approach. We generalise the approach to the wider class of \levy stochastic processes, both in the overdamped and in the underdamped regime, by finding exact equations for the relevant characteristic functions in Fourier space. 
\end{abstract}

\maketitle

\section{Introduction}

In a seminal 1926 paper, Richardson was able to show that, in the atmosphere, the average squared distance between two diffusing particles increases faster than linearly with time~\cite{richardson1926atmospheric}. This violation of Fick's law is due the turbulent nature of the atmosphere: in particular, Richardson observed that eddies tend to separate at a faster rate particles that are farther away from each other, and was able to determine the scaling $D\simeq l^{4/3}$ for the diffusivity, where $l$ is the distance between the particles. This is equivalent to saying that the mean square displacement of the particles is proportional to $t^3$ (unlike in the standard diffusion processes, where it is linear in time)~\cite{shlesinger1987levy}. Since then, anomalous diffusion has been recognized to be present in a wealth of domains in physics and beyond (e.g. in engineering, biology and finance)~\cite{Bouchaud1990AnomalousDI}, and many models have been proposed to describe and understand it~\cite{Hughes1982FractalRW,Metzler2014AnomalousDM}; among them, the class of \levy processes holds a prominent position~\cite{Metzler2000TheRW, dubkov2008}.

First introduced by Mandelbrot~\cite{mandelbrot1982fractal}, \levy flights are arguably the simplest realization of a super-diffusive stochastic process. They can be thought of as a sum of instantaneous displacements of a particle, following a \levy distribution; these jumps have the remarkable property that the sum of an arbitrary number of them is still a \levy random variable~\cite{Levy1955ThorieDL}. The name ``flights'' refers to the fact that these processes involve, from time to time, sudden fast displacements of the particle (the tails of the distribution are power-law like). Since these abrupt moves may reveal unphysical in many contexts, alternative descriptions based on the \levy statistics have been proposed: important examples are the truncated \levy flights~\cite{mantegna1994stochastic}, in which a suitable cutoff is imposed to the tails of the distribution, and the so-called \levy walks~\cite{zaburdaev2015levy}, in which the instantaneous velocity is bounded; in the latter case, the large displacements prescribed by the \levy statistics are achieved by keeping the same direction of motion for a suitable time. Still, pure \levy flights provide a useful model to study and understand phenomena subject to superdiffusive behaviour~\cite{palyulin2019first}, especially when used as the non-deterministic part of a Langevin-like equation (the so-called ``\levy noise'')~\cite{chechkin2002stationary}.

\levy processes have found applications in wide variety of fields, ranging from turbulence~\cite{shlesinger1986levy} to paleoclimate analysis~\cite{ditlevsen1999observation}, including finance~\cite{schoutens2003levy}. In condensed matter, they have been recognized to play an important role in Josephson junctions~\cite{augello2010non} and in the transport properties of disordered graphene~\cite{gattenlohner2016}. In plasma physics, it has been shown that the motion of the fast ions produced by nuclear fusion may be described by asymmetric \levy motion~\cite{bovet2014transport}. Also in biology, many observed behaviours can be characterized by using this class of stochastic models~\cite{reynolds2018current}. The interest around them  arose in the wake of the
influential paper by Viswanathan et al.~\cite{viswanathan1996levy}, observing \levy statistics in the foraging behaviour of wandering albatrosses. These results were later revisited, due to some methodological inconsistencies~\cite{edwards2007revisiting}, but they were nonetheless able to raise large interest in the biophysics community~\cite{reynolds2018current}, especially about the relation between optimal search strategies and \levy walks/flights~\cite{RevModPhys.83.81, viswanathan2011physics}. Nowadays non-Gaussian processes are observed also in completely different contexts, as in the path of eukariotic cells (whose motion is not determined by foraging~\cite{leptos2009dynamics}), swarming bacteria~\cite{ariel2017chaotic}, and cancer cells~\cite{huda2018levy}.

Due to the large number of potential applications, the behaviour of \levy processes subjected to external forces has been widely studied over the years~\cite{west1982linear,jespersen1999levy,chechkin2002stationary}. Particular attention has been devoted to understand to what steady states the particles relax, depending on the shape of the fixed external potential~\cite{chechkin2003bifurcation,ciesla2019multimodal}. From the point of view of practical applications, a further step would consist in understanding how the external potential needs to be manipulated, in order to bring the system to a desired final state in a finite time (and, possibly, in an optimal way). Let us consider, for instance, the situation in which a particle is subjected to an external harmonic confining potential, whose stiffness $k$ can be controlled in time. At the beginning the value of this elastic constant is $k_i$, and the particle is found in the corresponding stationary state. We want to bring it to the final steady state corresponding to $k=k_f$ in a given time $t_f$. If we just abruptly change the value of $k$, the relaxation of the system will take, in general, a time much longer than $t_f$; the time-dependent protocol $k(t)$ must be thus carefully chosen. Moreover, among the eligible protocols, it is interesting to search for that minimizing some cost function of the problem (as, for instance, the average work, the entropy production or the total time, given come constrains).

This class of problems, which are known under the name of ``shortcuts to adiabaticity'' (STA), is rooted in the context of quantum mechanics~\cite{torrontegui2013shortcuts}. The interest for them has then spread also in the domain of kinetic theory, with application to the study of Boltzmann equation~\cite{guery2014nonequilibrium}, and stochastic thermodynamics (see~\cite{guery2022driving} for a recent review). 
A successful approach to solve such problems is of inverse nature: one chooses a suitable time-dependent evolution
for the distribution of the quantity under study, from 
which the evolution equation allows to infer the time-dependent driving required. In general, several 
(infinitely many) types of driving are admissible, and a second level of question amounts to optimize some cost
function among the admissible family.
This method has been applied to many different systems, typically with the aim of switching between two different equilibrium states~\cite{martinez2016engineered, chupeau2018engineered}; recent studies have also addressed out-of-equilibrium problems, as the Brownian gyrator~\cite{baldassarri2020engineered} and driven granular gases~\cite{PhysRevResearch.3.023128, ruiz2022optimal}.


In this paper, we address the problem of finding STA for \levy processes driven by external harmonic potential.
The task is non-trivial in this case, because the stationary distributions associated to \levy processes are already hard to treat analytically. Yet, we need to go beyond stationarity, and find explicit time-dependent solutions.
The key ingredient, as we will show, is to consider the evolution of the characteristic function, which is more 
convenient to treat in this context.
 First, the overdamped limit is worked out in Section~\ref{sec:over};  it is possible in this case to find protocols corresponding to transformations in which the system is translated, and/or compressed (decompressed) by increasing (decreasing) the stiffness of the external controlling potential. In Section~\ref{sec:under}, we allow the particle to have inertia and we study the underdamped regime of the dynamics. 
 There, we are able to solve the problem for translation protocols. Conclusions are drawn in Section~\ref{sec:conclusions}

\section{Overdamped regime}
\label{sec:over}

As alluded to above, continuous stochastic processes ruled by \levy statistics are ubiquitous in physics. To characterise these dynamics it is useful to introduce a white stationary \levy noise, i.e. a stochastic process $\xi_{\alpha}(t)$ such that its integral over time
\begin{equation}
 I_{\alpha}(t) = \int_0^t dt'\,\xi_{\alpha}(t')
\end{equation} 
has stationary independent increments and characteristic function
\begin{equation}
\label{eq:charform}
 \widehat{p}_{I_{\alpha}}(s;t)=e^{-|s|^{\alpha}K_{\alpha}t}\,.
\end{equation}
We recall that the characteristic function $\widehat{p}(s)$ of a probability density function (PDF) $p(x)$ is defined as
\begin{equation}
 \widehat{p}(s)=\int_{-\infty}^{\infty} \,dx \, e^{isx}p(x)\,.
\end{equation}
Here, $\alpha \in (0,2]$ is the \levy index, and $K_{\alpha}$ is a constant with the physical dimensions of a length to the $\alpha$th power, divided by a time, which rules the intensity of the \levy noise. In the Brownian case $\alpha=2$, $I_2$ reduces to the usual Wiener process, and $K_2$ is the diffusion coefficient.
The $s \to -s$ symmetry of the characteristic function~\eqref{eq:charform} induces \textit{symmetric} \levy flights, meaning that displacements in the positive and in the negative direction covering the same distance are equally probable. Asymmetric noises are also possible, but they will not be considered in this paper. Appendix~\ref{sec:applevy} provides a minimal introduction to \levy $\alpha$-stable distributions.

In this Section, we will focus on the class of one-dimensional processes $x(t)$ whose dynamics can be modeled by a first-order stochastic differential equation of the form
\begin{equation}
\label{eq:sdeadditive}
 \dot{x}=\moby f(x) + \xi_{\alpha}(t)\,.
\end{equation} 
 The above dynamics can be seen as the overdamped motion of a particle subjected to the force $f(x)=-\partial_x U(x)$ deriving from an external potential $U(x)$, in a viscous medium with mobility $\moby$. The non-deterministic part of the evolution, $\xi_{\alpha}$, is a Lévy noise, with Lévy parameter $\alpha$ and generalized diffusion coefficient $K_{\alpha}$~\cite{chechkin2002stationary}.

It can be shown~\cite{ditlevsen1999anomalous,jespersen1999levy} that the PDF of the above processes obeys the Fractional Fokker-Planck equation  
\begin{equation}
\label{eq:ffpadditive}
 \partial_t p(x,t) = -\moby\partial_x[f(x)p(x,t)]+K_{\alpha}\frac{\partial^{\alpha} p(x,t)}{\partial |x|^{\alpha}}  \,,
\end{equation} 
where the Riesz fractional derivatives $\frac{d^{\alpha} }{d |x|^{\alpha}}$ are defined through their Fourier Transform
\begin{equation}
 \int_{-\infty}^{\infty}\,dx\,e^{-isx}\cbr{\frac{d^{\alpha} }{d |x|^{\alpha}} \varphi(x)}\, = - |s|^{\alpha}\int_{-\infty}^{\infty}\,dx\,e^{-isx}\varphi(x)\,.
\end{equation} 
It can be checked 
that if $\alpha=2$, the usual Fokker-Planck equation is recovered.

\subsection{Stationary solution in harmonic potential}

If the external potential is quadratic,
\begin{equation}
 U(x)=\frac{1}{2}\stiff (x-\mean)^2\,,
\end{equation}
where $\stiff$ is the stiffness and $\mean$ the rest position (point of zero force), then Eq.\eqref{eq:sdeadditive} reads
\begin{equation}
    \label{eq:langevinover}
\dot{x}=\moby k(\lambda-x) + \xi_{\alpha}(t)\,,
\end{equation}
while the fractional Fokker-Planck equation~\eqref{eq:ffpadditive} can be written as
\begin{equation}
\label{eq:ffpequaddim}
 \partial_t p = \moby \stiff\partial_x[(x-\mean)p]-\frac{K_{\alpha}}{2\pi} \int_{-\infty}^{\infty} \,e^{-isx}\,|s|^{\alpha}\,\widehat{p}(s,t)\,ds\,.
\end{equation}  

Starting from a given initial stationary state, we are concerned with the problem of finding protocols to reach a different
stationary state, in a prescribed time. To this end, the control we have over the system is through the
time-dependence of both the stiffness $k$ and the rest point $\lambda$.
The final state is completely specified by the values of the external potential parameters at the end of the process, namely
\begin{equation}
\begin{aligned}
\stiff(t_f)&=\stiff_f\\
\mean(t_f)&=\mean_f\,.
\end{aligned}
\end{equation}
If the external potential was suddenly switched into its final form, the typical time scale for the relaxation would be 
\begin{equation}
\tau=\frac{1}{\mu \stiff_f}\,.
\end{equation}

It is useful to turn to dimensionless units, through the change of variables
$$
t\to \tau \dimless{t} \quad \quad x\to \cbr{K_{\alpha}\tau}^{1/\alpha} \dimless{x} \quad \quad s\to \cbr{K_{\alpha}\tau}^{-1/\alpha} \dimless{s}\quad
$$
$$
\quad \mean\to \cbr{K_{\alpha}\tau}^{1/\alpha} \dimless{\mean} \quad \quad \moby \stiff \to \dimless{\stiff}/\tau\,.
$$
Eq.~\eqref{eq:ffpequaddim} can then be rewritten as
\begin{equation}
\label{eq:ffpequad}
\begin{aligned}
 \partial_{\dimless{t}} p = & \dimless{\stiff}\partial_{\dimless{x}}[(\dimless{x}-\dimless{\mean})p]\\
 &- \int_{-\infty}^{\infty}d\dimless{s} \,\frac{e^{-i\dimless{s}\dimless{x}}}{2\pi}\,|\dimless{s}|^{\alpha}\,\widehat{p}(\dimless{s},\dimless{t})\,.
 \end{aligned}
\end{equation} 
In these dimensionless variables one has, by definition, $\dimless{\stiff}(\dimless{t}_f)=1$, and the time-scale for the relaxation is unity.
In the following, stars will be dropped, in order to avoid clutter.

From the fractional Fokker-Planck Equation~\eqref{eq:ffpequad}, by passing to Fourier space, one obtains an equation for the characteristic function:
\begin{equation}
\label{eq:charevol}
 \partial_t\widehat{p}=- \stiff s \cbr{\partial_s \widehat{p}-i \mean\widehat{p}}-|s|^{\alpha}\widehat{p}\,,
\end{equation} 
whose stationary solution is
\begin{equation}
\label{eq:statchar}
  \widehat{p}_{st}(s)=\exp\cbr{is\mean-\frac{|s|^{\alpha}}{{\alpha}\stiff}};
\end{equation} 
the normalization condition $\widehat{p}_{st}(0)=1$ has been already taken into account.

To obtain the stationary distribution, we get back to real space:
\begin{equation}
\label{eq:statpdf}
 p_{st}(x)=\frac{1}{\pi}\int_0^{\infty}ds\,\cos(sx-s\mean)e^{-s^{\alpha}/{\alpha}\stiff}\,.
\end{equation} 
The above integral converges for all values $\alpha \in (0,2]$, but only for some of them is it possible to express the stationary PDF in closed form. Let us notice for instance that
    in the  Brownian case, $\alpha=2$, the PDF~\eqref{eq:statpdf} reads:
\begin{equation}
 p_{st}(x)=\sqrt{\frac{\stiff}{2\pi }}\exp\sbr{-\frac{\stiff}{2}(x-\mean)^2}\,,
\end{equation} 
which is consistent with the well known equilibrium distribution for a Brownian particle.
If $\alpha=1$ the solution is given instead by a Cauchy distribution~\cite{west1982linear}:
\begin{equation}
 p_{st}(x)=\frac{1}{\pi}\frac{\stiff}{1+\sbr{\stiff(x-\mean)}^2}\,.
\end{equation}

\subsection{Shortcuts to adiabaticity}
\label{sec:sta_overdamped}
Most STA protocols can be recast in the following procedure. Let us assume that we are interested in the stochastic process described by the evolution equation
\begin{equation}
\label{eq:genfp}
    \partial_t p(x,t)=\mathcal{F}[p](x,t;\{\zeta_i\})\,,
\end{equation}
where $\mathcal{F}[\cdot](x,t;\{\zeta_i\})$ is some evolution operator (e.g., the Fokker-Planck one) that depends on the set of control parameters $\{\zeta_i\}$. We need to find a suitable ansatz $p(x,t|\{\widetilde{\zeta}_i\})$ for the time-dependent solution, depending on the free parameters $\{\widetilde{\zeta}_i\}$, such that Eq.~\eqref{eq:genfp} reduces to a tractable system of equations relating $\{\zeta_i\}$ to $\{\widetilde{\zeta}_i\}$.
At this point the evolution of $\{\widetilde{\zeta}_i(t)\}$ can be chosen according to some criterion (e.g., optimization of a cost function during the process), and corresponding equations for the protocol $\{\zeta_i(t)\}$ are found in turn.

The same procedure could be adopted,  in principle, also in this case. To this end, working in Fourier space turns out to be more convenient when dealing with fractional values of $\alpha$. We therefore search for  time-dependent characteristic functions solving Eq.~\eqref{eq:charevol}. In this respect, the most natural ansatz for the solution is given by
\begin{equation}
\label{eq:ansatz}
\widehat{p}(s,t)=\exp\cbr{is\tmean(t)-\frac{|s|^{\alpha}}{\alpha \tstiff(t)}}\,,
\end{equation} 
where $\tmean(t)$ and $\tstiff(t)$ are time-dependent parameters whose evolutions still have to be fixed. Note that $\tmean$ is the median of the distribution: it can be checked that the distribution $p(x,t)$ stemming from Eq.~\eqref{eq:ansatz} is symmetric under $(x-\tmean) \to -(x-\tmean)$ transformations. For $\alpha > 1$, this quantity is also the mean value (which is not defined for $\alpha\le1$).

We insert the proposed solution~\eqref{eq:ansatz} into the evolution equation for the characteristic function, Eq.~\eqref{eq:charevol}, in order to get an explicit expression for $\stiff(t)$ and $\mean(t)$. The resulting condition reads
\begin{equation}
 i\dot{\tmean}s + \frac{|s|^{\alpha}}{{\alpha} \tstiff^2}\dot{\tstiff} = - i\stiff  \cbr{\tmean-\mean}s+(\stiff-\tstiff)\frac{|s|^{\alpha}}{ \tstiff s}\,.
\end{equation} 
By splitting the real and the imaginary part of the above equation, two coupled relations are found:
\begin{subequations}
\label{eq:solutions}
\begin{eqnarray}
\label{eq:solutions_stiff}
 \stiff &= \tstiff+\frac{\dot{\tstiff}}{ \alpha  \tstiff}\,,\\
\label{eq:solutions_mean}
 \mean &=\tmean+\frac{\dot{\tmean}}{ \stiff}\,.
\end{eqnarray} 
 \end{subequations}
The coupled equations~\eqref{eq:solutions} provide the time-dependent protocols $\stiff(t)$ and $\mean(t)$, once the evolution of the PDF is chosen (i.e., once $\tstiff(t)$ and $\tmean(t)$ are fixed). The driving protocol is thus inferred by first imposing the desired PDF evolution: let us stress that the success of this ``reverse engeneering'' technique relies on the possibility of finding a suitable ansatz for the time-dependent PDF, leading to conditions which are independent of $x$ (Eq.~\eqref{eq:solutions} in the present case).

The following boundary conditions need to be enforced:
\begin{equation}
\begin{aligned}
     \tmean(0)&=\mean(0)=\mean_i\quad\quad      & \tmean(t_f)&=\mean(t_f)=\mean_f  \\
 \tstiff(0)&=\stiff(0)=\stiff_i\quad\quad   & \tstiff(t_f)&=\stiff(t_f)=1\,.   
\end{aligned}
\end{equation}
The last condition follows from the adopted dimensionless units.

One way to determine the protocol is to assume that both $\tstiff(t)$ and $\tmean(t)$ are third-order polynomials. With this choice one finds
\begin{subequations}
\begin{eqnarray}
\label{eq:ktilde}
 \tstiff(t) &=\stiff_i+&\Delta \stiff (3 z^2-2 z^3)\\
\label{eq:ltilde}
\tmean(t) &=\mean_i+&\Delta \mean ( 3 z^2-2z^3)\,.
\end{eqnarray} 
\end{subequations}
where $\Delta \stiff = 1-\stiff_i$,  $\Delta \mean = \mean_f-\mean_i$ and we have introduced the rescaled time 
\begin{equation}
  z = t/t_f\,.
\end{equation}
Once inserted into Eq.~\eqref{eq:solutions}, the above expressions provide the explicit protocol we were looking for. In particular, the stiffness is described by
\begin{equation}
\label{eq:protstiff}
    \stiff=\stiff_i+\frac{1}{\alpha t_f} \frac{6 \Delta \stiff (1-z)z}{\stiff_i+\Delta \stiff (3-2z)z^2}+\Delta \stiff (3z^2 - 2 z^3)\,.
\end{equation}
If $\alpha=2$, the usual protocol for the overdamped Brownian case is recovered~\cite{martinez2016engineered}.
An analogous expression for the point of zero force is readily found:
\begin{equation}
\label{eq:protmean}
    \mean=\mean_i+\Delta \mean \frac{6 -2 \stiff t_f z^2 + 3 z ( \stiff t_f -2)}{ \stiff t_f}z\,.
\end{equation}

It is important to notice that the above derived relations provide protocols for arbitrary small values of $t_f$, while the spontaneous relaxation of the system would be observed, with the chosen dimensionless units, only on time-scales $t_f \gg 1$.

\subsection{Translation protocols}
Let us first focus on the particular case in which the stiffness is the same at the beginning and at the end of the process, and only the value of $\mean$ is required to change in time, corresponding to a mere translation.

\begin{figure}[ht]
    \centering
    \includegraphics[width=.9\linewidth]{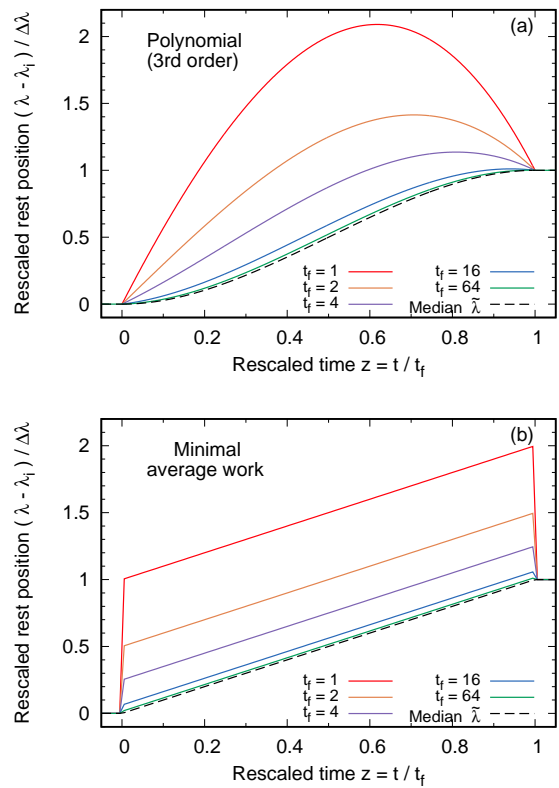}
    \caption{Translation protocols. Panel~(a): for different values of $t_f$, the time dependent protocol $\mean(t)$ defined by Eq.~\eqref{eq:protmean} is shown. The imposed evolution of $\tmean(t)$ [polynomial function in Eq.~\eqref{eq:ltilde}] is also displayed for comparison (dashed black curve). Panel~(b): the protocol that minimizes the average work, Eq.~\eqref{eq:optwprot}, is shown for different values of $t_f$. As before, the corresponding $\tmean(t)$ is reported as a dashed black line. The considered protocols do not depend on $\alpha$.}
    \label{fig:translation}
\end{figure}

If we require that the median $\tmean$ of the distribution follows the polynomial evolution defined by Eq.~\eqref{eq:ltilde}, the protocol to impose is given by Eq.~\eqref{eq:protmean}, with constant $\stiff=1$. As a consequence, the process does not depend on $\alpha$. This is a general property that comes from the fact that $\alpha$ does not appear in Eq.~\eqref{eq:solutions_mean}; for pure translational processes, the relations already known for the Brownian limit hold also for generic \levy distribution of the noise. It should be noticed that as soon as $\alpha>1$, the median $\tmean$ is also the average of the PDF, and Eq.~\eqref{eq:solutions_mean} can be derived by averaging the Langevin equation~\eqref{eq:langevinover} under the assumption of constant stiffness. The above described argument, making use of characteristic functions, is valid also for $\alpha \le 1$.

In Fig.~\ref{fig:translation}(a) the evolution of $\tmean$ is shown for different values of $t_f$.
  With our choice of the dimensionless units, the typical relaxation time of the dynamics is unity.
Consistently, the curves approach the quasi-stationary behaviour $\mean(t)=\tmean(t)$ when $t_f \gg 1$, since in this limit the ``thermalization'' of the system is much faster than the driving dynamics and $\tmean(t)$ closely ``follows'' the parameter $\mean(t)$: this slow driving regime corresponds to the ``adiabatic'' limit, to which the ``A" in ``STA" refers to. 
Conversely, when $t_f \simeq O(1)$, the protocol $\mean(t)$ can significantly differ from $\tmean(t)$.

The evolution~\eqref{eq:ltilde} is an arbitrary choice, and different functions can be taken, depending on the specific requirements of the problem under study. For instance, one may be
interested in minimizing the work needed, on average, to accomplish the protocol:
\begin{equation}
\begin{aligned}
 \av{W}&=\int_0^{t_f}dt\,\int_{-\infty}^{\infty}dx\, \partial_t U (x,t) p(x,t)\,. \\
 &=- \int_0^{t_f}dt\,\dot{\mean}\int_{-\infty}^{\infty}dx\,(x-\mean)p(x,t) \,.
\end{aligned}
\end{equation} 
The statistical properties of the \levy distributions assure that the above integral is well defined for $\alpha > 1$. For smaller values of $\alpha$ the average work diverges.

Taking into account the form of our ansatz~\eqref{eq:ansatz}, we can write this average work as
\begin{equation}
\begin{aligned}
 \av{W}&=-\frac{1}{2 \pi}\int_0^{t_f}dt\,\dot{\mean}\int_{-\infty}^{\infty}dx\,(x-\mean)\int_{-\infty}^{\infty}ds\,e^{is\cbr{x-\tmean}-|s|^{\alpha}/{\alpha}} \\
 &=\frac{i}{2 \pi}\int_0^{t_f}dt\,\dot{\mean}\int_{-\infty}^{\infty}ds\, e^{-|s|^{\alpha}/{\alpha }-is(\tmean-\mean)}\partial_s \int_{-\infty}^{\infty}dx\,e^{isx}\\
 &=i\int_0^{t_f}dt\,\dot{\mean}\int_{-\infty}^{\infty}ds\, e^{-|s|^{\alpha}/{\alpha }-is(\tmean-\mean)}\partial_s \delta(s)\,,
\end{aligned}
\end{equation} 
where in the first step we have applied the shift $x \to x + \mean$ to the integration variable, and then we have recognized the Fourier transform of a Dirac delta. By performing an integration by parts, under the proviso that $\alpha > 1$, we get 
\begin{equation}
\label{eq:avwork}
 \av{W}=-\int_0^{t_f}dt\, \dot{\mean}(\tmean-\mean)=\frac{\dot{\tmean}^2(t_f)-\dot{\tmean}^2(0)}{2}+\int_0^{t_f}dt \dot{\tmean}^2\,,
\end{equation} 
where  use was made of Eq.~\eqref{eq:solutions}.
The above integral is minimized by a motion with constant speed $\dot{\tmean}=\Delta \mean /t_f$, where $\Delta \mean = \mean_f - \mean_i$; indeed, the Euler-Lagrange equation reduces to $\ddot{\tmean}=0$, and the value of $\dot{\tmean}$ is fixed by the boundary conditions. The remaining terms on the right hand side of Eq.~\eqref{eq:avwork} vanish in the present case (as we demand for steady states at $t=0$ and $t=t_f$).
The evolution of $\tmean$ and the corresponding protocol for the rest position $\mean$ of the external potential  then read
\begin{subequations}
\begin{equation}
\label{eq:optwevo}
 \tmean=\tmean_i + \Delta \mean\, z
\end{equation}
\begin{equation}
\label{eq:optwprot}
 \mean=\tmean_i + \Delta \mean \cbr{z+\frac{1}{ t
 _f}}\,.
\end{equation} 
\end{subequations}
It is worth noticing that in order to fulfill the boundary conditions, sudden jumps are needed to the value of $\mean$ at the beginning and at the end of the process, in agreement with previous works pertaining to the Brownian case~\cite{schmiedl2007optimal}. These discontinuities have no consequence on the average work, which can be written as a function of the time evolution of $\tmean$ only [see Eq.~\eqref{eq:avwork}].
Figure~\ref{fig:translation}(b)
presents the situation, where  the curve of $\mean$ again approaches that of $\tmean$ (quasi-static limit) as $t_f \gg 1$.

\subsection{Compression/decompression protocols}
Another particular case of the protocols described in Section~\ref{sec:sta_overdamped} is met when the rest position of the external potential does not change during the process, and only the stiffness $\stiff$ is varied. Depending on the sign of $\Delta \stiff= 1 - \stiff_i$, one then achieves a ``compression'' or a ``decompression'' (we recall that with our choice of the dimensionless units, $\stiff(t_f)=1$).

\begin{figure}[ht]
    \centering
    \includegraphics[width=.9\linewidth]{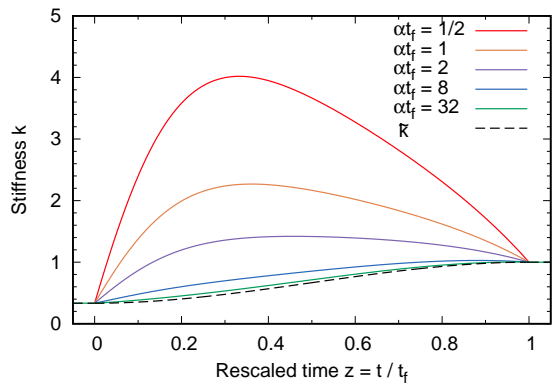}
    \caption{Compression protocols. The evolution of $\stiff$ that is required to compress the system from $\stiff_i=1/3$ to $\stiff_f=1$ is  shown for several values of $\alpha t_f$. The imposed evolution for $\tstiff$ [the 3rd order polynomial Eq.~\eqref{eq:ktilde}] is represented as a black dashed curve.  For $\alpha t_f \gg 1$ the evolution approaches the quasi-stationary limit $\stiff(z) \simeq \tstiff(z).$}
    \label{fig:compression}
\end{figure}

\begin{figure}[ht]
    \centering
    \includegraphics[width=.9\linewidth]{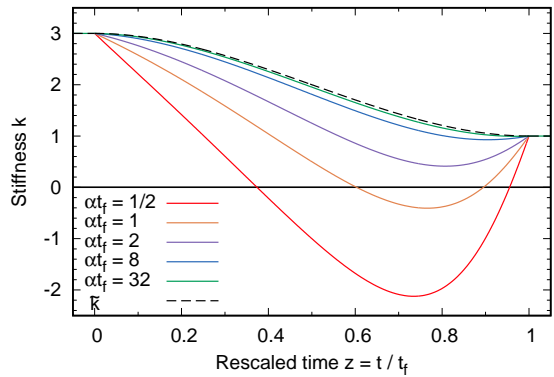}
    \caption{Decompression protocols. Evolution of $\stiff$ when decompressing the system from $\stiff_i=3$ to $\stiff_f=1$, according to Eq.~\eqref{eq:ktilde}, for different values of $\alpha t_f$.
    Notice that some of the curves involve negative values of the stiffness to impose. Also for the decompression protocols, the quasi-static limit is approached as $\alpha t_f \gg 1$.}
    \label{fig:decompression}
\end{figure}

In Fig.~\ref{fig:compression} and~\ref{fig:decompression}, different drivings as encoded in Eq.~\eqref{eq:protstiff} are shown, for both compression and decompression. For increasing values of $\alpha t_f$, as expected, the protocols approach the imposed $\tstiff(z)$, determined in this case by Eq.~\eqref{eq:ktilde}.

Unlike translations, (de)compression protocols do depend on the \levy parameter $\alpha$. Once expressed in terms of the rescaled time $z$, the evolution of $\stiff$ is a function of the product $\alpha t_f$: as a consequence, for decreasing values of $\alpha$ the curves will move away from the imposed $\tstiff(z)$ evolution (which is expected to be equal to $\stiff(z)$ in the opposite, quasi-static limit, $\alpha t_f \to \infty$). This can be understood by looking at Figs.~\ref{fig:compression} and ~\ref{fig:decompression}, where the value of $\alpha t_f$ is changed. In particular, if the transition is required to happen in a rather short time interval $t_f$, a decompression protocol may involve negative values of $\stiff$. This condition is fine from a mathematical point of view, but it means that the trap should be transiently expulsive rather than confining, which 
may lead to practical difficulties in applications~\cite{Bayati_2021}. It is thus natural to wonder what condition must be imposed on the parameters of the problem in order to keep positive values of $\stiff$ during the whole decompression process. Multiplying Eq.~\eqref{eq:solutions}(a) by $\alpha /\tstiff$ leads to the relation
\begin{equation}
    \alpha  + \frac{\dot{\tstiff}}{\tstiff^2}=\alpha \frac{\stiff}{\tstiff } \ge 0
\end{equation}
where the inequality holds if the external stiffness is constrained to non-negative values.
By integrating between $t=0$ and $t=t_f$ one gets
\begin{equation}
 \alpha t_f \ge \frac{1}{\stiff_f}-\frac{1}{\stiff_i}\,.
\end{equation} 
The equality holds when the external potential is suddenly removed at the beginning of the process and then restored at the end, so that during the time interval $\stiff=0$ the evolution is completely free.

Leaving aside the particular case $\alpha=2$, the average work is not well-defined along a (de)compression protocol. Indeed, to evaluate that quantity one should compute the integral
\begin{equation}
\begin{aligned}
 \av{W}= \int_0^{t_f}dt\,\dot{\stiff}\int_{-\infty}^{\infty}dx\,(x-\mean)^2p(x,t) \,,
\end{aligned}
\end{equation} 
which is ill-defined for $\alpha<2$. As a consequence, in this case it is meaningless to search for the protocol which minimizes the work. For the Brownian case, the problem has been studied in several works~\cite{schmiedl2007optimal,schmiedl2007efficiency,PhysRevLett.106.250601,plata2019optimal}.

\subsection{Compound protocols}

Enforcing a simultaneous translation and (de)compression may lead to quite involved dynamics, due to the coupling between $\mean$ and $\stiff$ in Eq.~\eqref{eq:solutions_mean}.
\begin{figure}
    \centering
    \includegraphics[width=.9\linewidth]{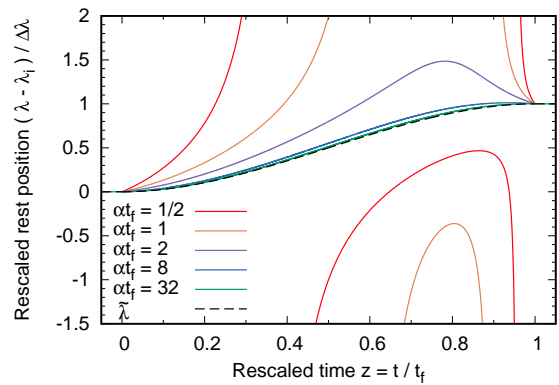}
    \caption{Compound protocols involving simultaneous translation and decompression (overdamped dynamics). The rest position $\mean$ of the external potential (point of zero force) is shown, for a compound protocol in which the stiffness decreases from $\stiff_i=3$ to $\stiff_f=1$.
    The curves follow Eq.~\eqref{eq:protmean}, while the evolution of $\stiff$ (not shown) is the same as in Fig.~\ref{fig:decompression} [computed from Eq.~\eqref{eq:protstiff}]. Different values of $t_f$ are considered: as before, the curve approaches the imposed $\tmean$ evolution in the quasi-static limit $\alpha t_f \to \infty$.}
    \label{fig:compound}
\end{figure}

\begin{figure*}[htbp]
    \centering
    \includegraphics[width=.9\linewidth]{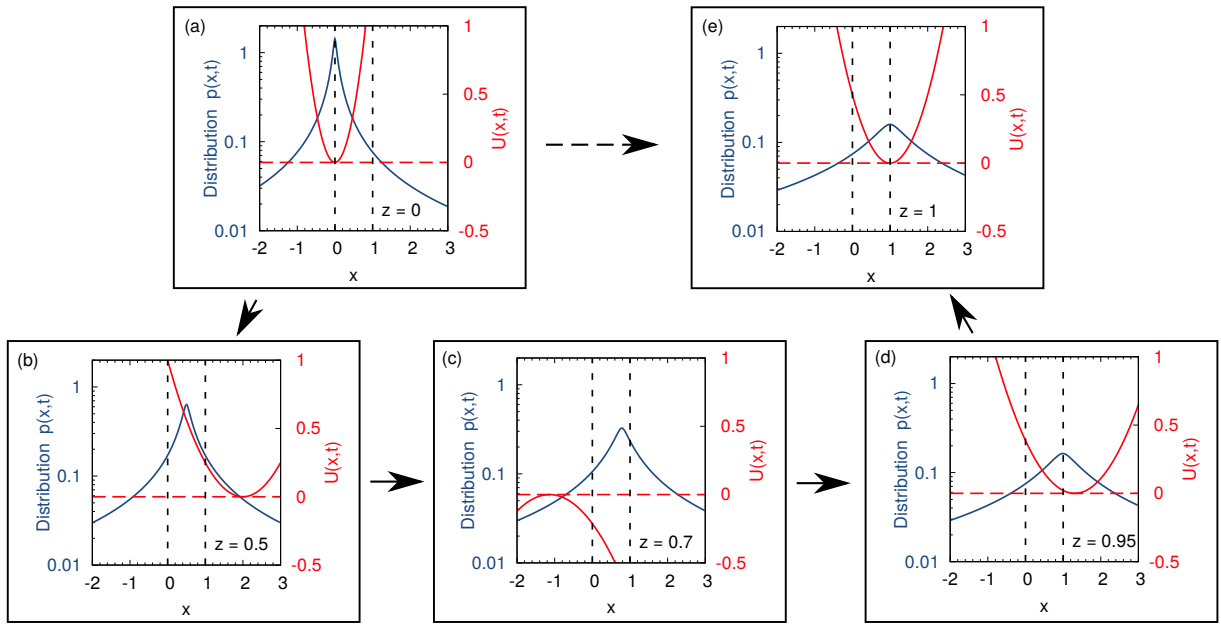}
    \caption{Pushing-pulling protocol (overdamped dynamics). Time evolution of distribution function and external potential in a compound  translation-decompression protocol leading the system from the initial steady state in panel (a) to the final steady state in panel (e). The instantaneous distribution (blue) and the external driving potential (red) are shown at different stages of the protocol. Following the evolution of the system, we note that at intermediate times [panel (c), z=0.7] the stiffness of external potential changes sign and the rest point switches from positive to negative, before reaching the final configuration. In other words, in panels (b) and (d) the  potential ``pulls'' the distribution; in panel (c), on the other hand, the particle is ``pushed'' by the external force.  Parameters:  $\alpha=1/2$, $\stiff_i=3$, $\stiff_f=1$, $\mean_i=0$, $\mean_f=1$, $t_f=2$, corresponding to the protocol shown in Fig.~\ref{fig:compound}, yellow curve.}
    \label{fig:movie}
\end{figure*}

Some examples are provided in Fig.~\ref{fig:compound}, where the rest position Eq.~\eqref{eq:protmean} is computed for different compound translation-decompression protocols. When $t_f$ is small enough, negative values of $\stiff$ are induced, as shown in Fig.~\ref{fig:decompression}(b). When $\stiff$ becomes equal to zero, due to Eq.~\eqref{eq:protmean}, $\mean$ tends to $\infty$. At that point the external potential is flat, and the particle is (momentarily) free. Moreover, as $\stiff$ becomes negative, $\lambda$ changes sign too, passing from $+\infty$ to $-\infty$: not only the curvature of the external potential is reversed, but also the point of zero force is on the other side of the real axis, with respect to the median $\tmean$ of the distribution. In some sense, the external force, which at the beginning of the process was ``pulling'' the particle, is now ``pushing'' it. The situation is reversed again when $\stiff$ turns back to positive values before reaching its final value $\stiff_f$. A pictorial representation of the process is provided in Fig.~\ref{fig:movie}, where the external potential and the distribution are plotted at different times.

\section{Underdamped dynamics}
\label{sec:under}
Let us now consider the underdamped version of the model described in Section~\ref{sec:over}, i.e. the case of a particle with inertia subject to \levy noise in an harmonic trap. The motion is described by the equations
\begin{equation}
\label{eq:genkk}
 \begin{cases}
  \dot{x}=v\\
  m\dot{v}=-\partial_xU(x)-\frict v+\xi_{\alpha}(t)
 \end{cases}
\end{equation}
where the \levy noise $\xi_{\alpha}$ features the same properties as discussed for the overdamped case. Here $v$ is the instantaneous velocity of the particle, $m$ is the mass and $\frict=1/\moby$ the damping coefficient. Equations~\eqref{eq:genkk} tend to the Klein-Kramers description for the special case $\alpha=2$~\cite{risken}. The above evolution can be written in terms of a second-order stochastic differential equation for the position as
\begin{equation}
\label{eq:langevininertia}
    m\ddot{x}=-\partial_xU(x)-\frict \dot{x}+\xi_{\alpha}(t)\,,
\end{equation}
or, equivalently, as the
fractional Fokker-Planck equation~\cite{west1982linear,lu2011inertial}
\begin{equation}
 \partial_t p = - \partial_x (v p) + \frac{1}{m}\partial_v \sbr{\partial_xU p + \frict v p }+K_{\alpha}\frac{\partial^{\alpha}p}{\partial |v|^{\alpha}}\,.
\end{equation} 

\subsection{Stationary solution in harmonic potential}

We now specialize to the harmonic case 
\begin{equation}
 U(x)=\frac{\stud}{2}(x-\mean)^2
\end{equation} 
and, as before, we switch to dimensionless variables 
$$
t\to \tau' \dimless{t} \quad \quad x\to \cbr{K_{\alpha}}^{1/\alpha}(\tau')^{1+\frac{1}{\alpha}} \dimless{x} \quad \quad v\to \cbr{K_{\alpha}\tau'}^{1/\alpha} \dimless{v} \quad \quad
$$
$$\mean\to \cbr{K_{\alpha}}^{1/\alpha}(\tau')^{1+\frac{1}{\alpha}} \dimless{\mean} \quad \quad  \stud \to \frac{m}{(\tau')^2}\dimless{\stud}\,,
$$
where
$$
\tau'=\frac{m}{\frict}
$$
is the typical relaxation time-scale of the underdamped dynamics (decorrelation time of the velocity in the absence of external forces). Let us notice that in the underdamped regime $\tau'$ is larger than $\tau= \frict/k_f$ (the relevant time-scale for the overdamped case). The geometrical average $\sqrt{\tau \tau'}$ is proportional to the characteristic period of the harmonic oscillator.

Dropping the stars, the fractional Fokker-Planck equation in the new variables reads
\begin{equation}
  \partial_t p = - \partial_x (v p) + \partial_v \sbr{\stud(x-\lambda) p + v p }+\frac{\partial^{\alpha}p}{\partial |v|^{\alpha}}\,.
\end{equation} 
It is useful to define the typical angular frequency of the damped oscillator
\begin{equation}
 \omega=\sqrt{\stud-\frac{1}{4}}\,.
\end{equation}
Since we are interested in the underdamped limit, we assume that the argument of the square root is positive, and $\omega$ is thus real.

We introduce the characteristic function
\begin{equation}
 \widehat{p}(s,u,t)=\int_{-\infty}^{\infty}dx\int_{-\infty}^{\infty}dv\, e^{isx+iuv}p(x,v,t)\,,
\end{equation} 
so that the fractional Fokker-Planck equation can be rewritten as
\begin{equation}
\label{eq:undffp}
 \partial_t \widehat{p}= \cbr{s-u}\partial_u \widehat{p}-\stud u\partial_s \widehat{p} + i \stud \mean u \widehat{p} - |u|^{\alpha} \widehat{p}\,.
\end{equation} 
The stationary solution of the above equation can be found by means of the method of characteristics~\cite{west1982linear,lu2011inertial}. An explicit derivation is detailed in Appendix~\ref{sec:derivstat}. The final result is:
\begin{equation}
 \widehat{p}_{st}(s,u)=\exp\sbr{i \mean s   +  \frac{|u|^{\alpha}}{[g(y)]^{\alpha}}\int_0^y dy'[g(y')]^{\alpha} -\frac{ |s_0|^{\alpha}}{\alpha \stud}   }
\end{equation} 
where

\begin{subequations}
\begin{equation}
\label{eq:defy}
 y=y(s,u)=\frac{1}{\omega}\arctan\cbr{\frac{\omega}{\frac{s}{u}-\frac{1}{2}}}\,,
\end{equation} 
\begin{equation}
\label{eq:defs0}
    s_0=s_0(s,u)=\frac{\omega u}{g(y(s,u))}\,,
\end{equation}
\end{subequations}
and
\begin{equation}
 g(y)=\sin(\omega y)e^{- y/2 }\,.
\end{equation}

\subsection{STA for translation processes}

In this section we aim at finding explicit protocols to connect steady states with different values of $\mean$ (but same $\stud=\stud(t_f)=1$) in a finite time $t_f$.

As in the overdamped case, we need to assume a suitable ansatz for the shape of the characteristic function during the protocol; plugging it into Eq.~\eqref{eq:undffp} will provide a relation between the external control parameter $\mean$ and the time-dependent variables determining the shape of the pdf during the process.

Our ansatz reads
\begin{equation}
\label{eq:udansatz}
 \widehat{p}(s,u)=\exp\sbr{i \tmean(t) s   +  |u|^{\alpha}\mathcal{G}_{\alpha}(y) -\frac{|s_0|^{\alpha}}{\stud \alpha} + u h(t) }
\end{equation}
with $y=y(s,u)$ and $s_0=s_0(s,u)$ as defined in Eqs.~\eqref{eq:defy} and~\eqref{eq:defs0}, and
\begin{equation}
    \mathcal{G}_{a}(y)=\frac{1}{[g(y)]^{a}}\int_0^y dy'[g(y')]^a\,,
\end{equation}
where $h(t)$ is a time-dependent function such that $h(0)=h(t_f)=0$\,.
Exploiting linearity, Eq.~\eqref{eq:undffp} can be written in the more convenient form
\begin{equation}
\label{eq:logundffp}
 \partial_t \ln \widehat{p}= L\ln \widehat{p} + i\stud \mean u  - |u|^{\alpha} \,,
\end{equation} 
where we have introduced the linear operator
\begin{equation}
\label{eq:linearopl}
 L=   \cbr{s-u}\partial_u -\stud u\partial_s\,.
\end{equation}
In Appendix~\ref{sec:appprop} it is shown that, given a generic function $f(s_0)$,
\begin{equation}
\label{eq:propl1}
    L [f(s_0)]=0\,;
\end{equation}
moreover, from the property
\begin{equation}
\label{eq:propl2}
    L [u^a\mathcal{G}_a(y(s,u))]=u^a\,
\end{equation}
(shown again in Appendix~\ref{sec:appprop}), it can be concluded, by invoking the linearity of $L$, that
\begin{equation}
    L [|u|^{\alpha}\mathcal{G}_a(y(s,u))]=|u|^{\alpha}\,.
\end{equation}
Taking into account these results and our choice of the ansatz, Eq.~\eqref{eq:logundffp} leads to
\begin{equation}
\begin{aligned}
    i\dot{\tmean}s+u\dot{h}&=i\tmean Ls +h L u + i \stud \mean u\\
    &=-i\stud \tmean u + h\cbr{s-u} +i\stud \mean u\,.
\end{aligned}
\end{equation}
We require that the above equation holds for any value of $s$ and $u$; it follows that
\begin{subequations}
\label{eq:condundh}
\begin{equation}
h=i \dot{\tmean}    
\end{equation}
\begin{equation}
\label{eq:condund}
\mean=\tmean+\frac{ \dot{\tmean}+\ddot{\tmean}}{k}\,.
\end{equation}
\end{subequations}

 This formula provides the relation between $\mean$ and $\tmean$ we were searching for. The inertial term of the underdamped regime results in the appearance of the second order derivative of ${\tmean}$ in Eq.~(\ref{eq:condund}). As for the corresponding overdamped case, the protocol does not depend on the \levy index $\alpha$. In particular, it has to be the same also for the Brownian case $\alpha=2$; this verification is worked out in Appendix~\ref{sec:appundbrown}.

Let us notice that the validity of the relations~\eqref{eq:propl1} and~\eqref{eq:propl2} relies on the hypothesis that the values of $\stud$ in the ansatz and in the operator $L$ are the same; (de)compression processes with a distribution parameter $\tstud$ different from $\stud$, as in the overdamped case, would require more elaborated strategies.

Equation~\eqref{eq:condund} can be inferred from the very beginning by formally averaging Eq.~\eqref{eq:langevininertia} as a relation for the mean. 
This is how an identical relation is found, for instance, in~\cite{gomez2008optimal}, where a related problem, in the Brownian limit, is addressed. 
However, it should be kept in mind that for $\alpha \le 1$ the parameter $\tmean$ is \textit{not} the average of the distribution, which is actually not defined.


Once $\tmean(t)$ is fixed in such a way that the final state is reached in a time interval $t_f$, Eq.~\eqref{eq:condund} allows to compute the explicit expression for the external potential. In the same spirit of what has been done for the overdamped dynamics, also in this case we search for the simplest protocol fulfilling the boundary conditions
\begin{equation}
\begin{aligned}
     \tmean(0)&=\mean(0)=\mean_i\quad\quad      & \tmean(t_f)&=\mean(t_f)=\mean_f   
\end{aligned}
\end{equation}
and the constraint given by Eq.~\eqref{eq:condund}.
Since $\mean$ also depends on $\ddot{\tmean}$, in this case we need to impose
\begin{equation}
    \dot{\tmean}(0)=\dot{\tmean}(t_f)=0\,,
\end{equation}
to avoid discontinuities of $\dot{\tmean}$ at $t=0$ or $t=t_f$. Indeed, if $\dot{\tmean}\ne 0$ at the boundaries, due to Eq.~\eqref{eq:condund}, also $\ddot{\tmean}$ would be finite, leading to infinite instantaneous variation of the driving parameter $\mean$.

A relatively simple polynomial fulfilling all the above conditions is
\begin{equation}
\label{eq:lund}
    \tmean=\mean_i+\Delta \mean\, z^3\cbr{6z^2-15z+10}\,,
\end{equation}
leading to the external protocol
\begin{equation}
\label{eq:ltildeund}
\begin{aligned}
\mean=&\mean_i +  \Delta \mean\, z^3 \cbr{6 z^2-15 z +10 }+\\
&+30\frac{\Delta \mean}{ \stud t_f^2} z(z-1)\sbr{t_fz^2 + (4- t_f)z-2}\,.
\end{aligned}
\end{equation}

\begin{figure}
    \centering
    \includegraphics[width=.9\linewidth]{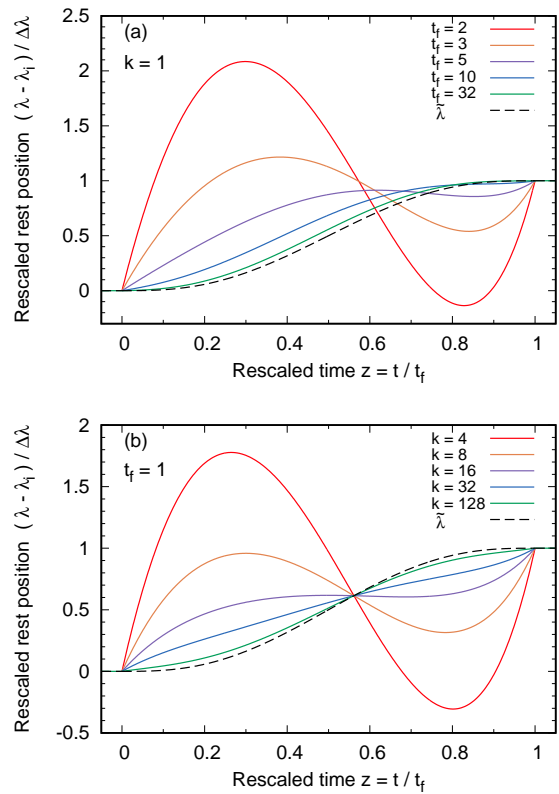}
    \caption{Translation protocol for underdamped dynamics. In both panels, the dashed line represents the imposed evolution $\tmean(z)$ of the median, Eq.~\eqref{eq:lund}, while the solid coloured curves are examples of protocol~\eqref{eq:ltildeund} for the rest position $\mean(z)$ of the external potential. In Panel~(a) different values of $t_f$ are considered, while the stiffness is fixed to $\stud=1$. Panel~(b) shows instead examples of $\mean(z)$ for a fixed value of $t_f=1$ and different choices of $\stud$. In the limits $t_f \to \infty$ and $\stud \to \infty$ the evolution approaches the quasi-static protocol $\mean(z)\simeq \tmean(z)$, as expected from Eq.~\eqref{eq:condund}.  As already seen for the overdamped case, the protocol does not depend on the \levy index $\alpha$.}
    \label{fig:undtransl}
\end{figure}

Figures~\ref{fig:undtransl}(a) and~\ref{fig:undtransl}(b) show the driving~\eqref{eq:ltildeund} for different values of $t_f$ and $\stud$, once the evolution~\eqref{eq:lund} has been imposed.
The quasi-static behaviour $\mean(t) \simeq \tmean(t)$ is approached in the limits $t_f \gg 1$ and $\stud \gg 1$. This can be expected on physical grounds, as both conditions imply that the typical time scales of the dynamics are much shorter than the total time of the protocol. It can be checked that these considerations are consistent with~Eq.~\eqref{eq:condund}.  We recall that, with the chosen dimensionless variables, spontaneous relaxation would be complete for $t_f \gg 1$.

ba

\subsection{Work optimization for translation processes}

It is interesting to look for the protocol which minimizes the average work in the underdamped case. For $\alpha > 1$ one has 
\begin{equation}
    \begin{aligned}
        \av{W}&=\int_{-\infty}^{\infty}\,dx\int_{0}^{t_f}\,dt\,\partial_t U(x,t)p(x,t)\\
        &=\frac{1}{2\pi}\int_{-\infty}^{\infty}\,dx\int_{0}^{t_f}\,dt\,\partial_t U(x,t)\int_{-\infty}^{\infty}ds\,\widehat{p}(s,0,t)\,.
    \end{aligned}
\end{equation}
        Recalling (see Appendix~\ref{sec:derivstat}) that $s_0 \to s$ for $u \to 0$ one has
       \begin{equation}
    \begin{aligned}
        \av{W} &=\int_{0}^{t_f} dt\,\frac{ \dot{\mean}}{2 \pi }\int_{-\infty}^{\infty}dx\,(\mean-x)\int_{-\infty}^{\infty}ds\,e^{-isx+i\tmean s -\frac{|s|^{\alpha}}{\alpha}}\\
     &=\int_{0}^{t_f} dt\,\frac{\dot{\mean}}{2 \pi  i }\int_{-\infty}^{\infty}ds\,e^{i\tmean s -\frac{|s|^{\alpha}}{\alpha}-i \mean s}\partial_s \delta(s)\\
     &=\int_{0}^{t_f} dt\, \dot{\mean}\cbr{\mean- \tmean }\,,
    \end{aligned}
\end{equation}
where first we have recognized the Fourier transform of a Dirac delta, and then we have integrated by parts.

Bearing in mind condition~\eqref{eq:condund} one finally has
\begin{equation}
\label{eq:avworkund}
    \av{W}=\int_0^{t_f}dt\, \cbr{\dot{\tmean}+\frac{\ddot{\tmean} +\dddot{\tmean}}{k} }\cbr{\dot{\tmean}+\ddot{\tmean}}\,.
\end{equation}

The evolution that minimizes $\av{W}$ is the one solving the Euler-Lagrange equation
\begin{equation}
    \partial_{\tmean} \mathcal{L}-\frac{d}{dt}\partial_{\dot{\tmean}}+\frac{d^2}{dt^2}\partial_{\ddot{\tmean}} \mathcal{L}-\frac{d^3}{dt^3}\partial_{\dddot{\tmean}} \mathcal{L}=0
\end{equation}
with
\begin{equation}
    \mathcal{L}(t, \tmean, \dot{\tmean}, \ddot{\tmean}, \dddot{\tmean})=\cbr{\dot{\tmean}+\frac{\ddot{\tmean} +\dddot{\tmean}}{k} }\cbr{\dot{\tmean}+\ddot{\tmean}}\,.
\end{equation}
The solutions are given by
\begin{equation}
    \ddot{\tmean}=0\,,
\end{equation}
which implies, accounting for the boundary conditions,
\begin{subequations}
\begin{equation}
    \tmean=\mean_i + \Delta \mean\, z\,,
\end{equation}
\begin{equation}
\label{eq:deltaprot}
    \mean=\mean_i + \Delta \mean\, z+ \frac{\Delta \mean}{ t_f} + \frac{ \Delta \mean}{ t_f^2}\sbr{\delta(z)-\delta(z-1)}\,.
\end{equation}
\end{subequations}
The protocol which minimizes the average work is thus quite similar to the one already seen for the overdamped case: it amounts to a rigid translation at constant speed of the distribution, obtained by ``dragging'' it through a linear motion of the external potential.
An important difference between the two situations lies though in the fact that here the discontinuities of $\dot{\tmean}$ at the boundaries lead to the presence of two delta-shaped terms. At the beginning of the protocol, an instantaneous ``kick'' is needed to increase the velocity of the translating distribution, while a sudden slowdown has to be imposed at the end. The qualitative scenario resembles the one found in~\cite{gomez2008optimal}, where a similar problem, in the Brownian case, is treated; in that context, however, the final value $\mean(t_f)$ is imposed instead of $\tmean(t_f)$, a difference which explains the discrepancy between the results found there and Eq.~\eqref{eq:deltaprot}. This means that in \cite{gomez2008optimal}, there is no
control on the final state reached, since the target 
pertains to the confining potential, not to 
the distribution of position and velocity. Also in this case, as in the overdamped situation, it should be noticed that the sudden jumps on $\mean$ do not affect the average work; indeed, $\av{W}$ can be written as a function of the time derivatives of $\tmean$ only, through Eq.~\eqref{eq:avworkund}.

\section{Conclusions}
\label{sec:conclusions}
\levy processes are a useful generalization of Brownian motion, able to describe a large gamut of stochastic dynamics in physics and beyond. We discussed how the problem of adiabaticity shortcuts generalises in this context. We have analyzed the case of a particle subject to \levy noise and harmonic confining potential, both in the overdamped and in the generic underdamped regime. In the former limit, we can find explicit analytical protocols for translation processes, (de)compressions and compositions of the two effects; in the latter, we have studied pure translations only.

In the Brownian case, the relations defining the external dynamical protocol can be typically found by analyzing the Fokker-Planck equation in real space; here, due to the peculiarities of \levy noise, an exact analysis is only possible in Fourier space, by making suitable ansatzs for the characteristic function. The two approaches coincide when the \levy stability parameter $\alpha$ is equal to 2 (Gaussian limit).

Once analytical relations for the protocols are available, it is also possible to optimize quantities of interest along the evolution. Here, we have considered the problem of optimal average work in translation processes, generalizing the results already known for the Brownian limit.

Along the lines of the present results, one may study the more involved case of underdamped processes with \levy noise and varying stiffness. Besides, our study shows that it is possible to apply the methods of shortcuts to adiabaticity to models whose stochastic nature is not described by the usual additive Gaussian noise; this opens a promising perspective on a wide class of out-of-equilibrium systems.

\appendix

\section{Basic properties of $\alpha$-stable \levy  distributions}
\label{sec:applevy}
A full discussion about \levy $\alpha$-stable distributions is beyond the scope of this paper. While referring the reader to specialized textbooks~\cite{Levy1955ThorieDL, zolotarev1986one, samorodnitsky2017stable},  we limit  ourselves here to an outline of their main properties.

A probability distribution $p$ is said to be stable if, given two random variables $x$ and $y$ such that
\begin{equation}
    x \sim p(x) \quad \quad    y \sim p(y)
\end{equation}
(here and in the following the symbol ``$\sim$'' means ``is distributed according to''), then any linear combination $z=ax+by$ of the two (with $a$ and $b$ real constants) satisfies
\begin{equation}
    z \sim p(cz +d)
\end{equation}
for some choice of $c$ and $d$.
 The most important example is the Gaussian, which is the only one with finite variance, and also one of the few that can be written in closed form.

In general, stable distributions can only be expressed by means of their characteristic function, i.e.
\begin{equation}
    \widehat{p}(s)=\int_{-\infty}^{\infty}ds\,e^{ixs}p(x)\,.
\end{equation}
It can be shown that all (and only) the distributions whose characteristic function reads
\begin{equation}
    \widehat{p}(s;\alpha, \beta, \gamma,\delta)=e^{is \delta - |\gamma s|^{\alpha}\cbr{1-i\beta \frac{s}{|s|}\phi(s)}}
\end{equation}
with 
\begin{equation}
    \phi(s)=\begin{cases}
 \cbr{|\gamma s|^{1-\alpha}-1}\tan\cbr{\frac{\pi \alpha}{2}}\\
 -\frac{2}{\pi} \log|\gamma s|
\end{cases}
\end{equation}
are stable. The parameter
$\alpha\in (0,2]$ is sometimes called ``\levy index''~\cite{chechkin2002stationary}; the Gaussian case is recovered when $\alpha=2$. The symmetry of the distribution is ruled by $\beta$ (it is symmetric if $\beta=0$).

\levy $\alpha$-stable distributions are known to have ``heavy tails'', meaning that their asymptotic behaviour (for $\alpha<2$) is power-law. In particular, it can be shown that
\begin{equation}
    p(x)\approx |x|^{-(1+\alpha)}\quad \text{when}\quad |x|\gg 1\,.
\end{equation}

A consequence of the stability property is that any random variable resulting from a sum process (i.e., an iterated sum of identically distributed random variables) will be described by a distribution belonging to this class. A generalized Central Limit Theorem holds~\cite{kolmogorov1968limit}.

\section{Stationary state for the underdamped harmonic oscillator with \levy noise}
\label{sec:derivstat}
To find the stationary solution for the underdamped harmonic oscillator in the case of generic \levy noise, we have to impose $\partial_t \widehat{p}=0$ in Eq.~\eqref{eq:undffp}.
The resulting equation for the steady state characteristic function,
\begin{equation}
\label{eq:statchar_app}
    \cbr{s-u}\partial_u \widehat{p}-\stud u\partial_s \widehat{p} + \cbr{i \stud \mean u - |u|^{\alpha}} \widehat{p}=0\,,
\end{equation}
is a linear partial differential equation which can be solved with the method of characteristics. It is worth recalling that here the term ``characteristics'' refers to a particular set of curves $f(s,u)=const$ in the $(s,u)$ plane, such that Eq.~\eqref{eq:statchar_app} becomes an ordinary differential equation when evaluated along any of those curves. They should not be confused with the characteristic functions of probability theory,
a terminology also used in the present paper.

We introduce a parametric description of the variables $s$, $u$ 
\begin{equation}
\begin{aligned}
s&=s(y)\quad\quad u=u(y)
\end{aligned}
\end{equation}
such that
\begin{equation}
    dy=\frac{du}{s- u}=-\frac{1}{\stud u}ds
\end{equation}  
or, equivalently,
\begin{equation}
\label{eq:odesu}
\begin{aligned}
    \frac{du}{dy}&=s - u\\
     \frac{ds}{dy}&= - \stud u\,.
\end{aligned}
\end{equation}
With this choice, Eq.~\eqref{eq:statchar_app} can be rewritten as
\begin{equation}
\label{eq:odechar}
    \frac{d \widehat{p}}{dy}=\frac{du}{dy}\partial_u\widehat{p}+\frac{ds}{dy}\partial_s\widehat{p}=-\cbr{i \stud \mean u - |u|^{\alpha}}\widehat{p}\,,
\end{equation}
i.e. an ordinary differential equation, much simpler to solve.

\begin{figure}[ht]
    \centering
    \includegraphics[width=.9\linewidth]{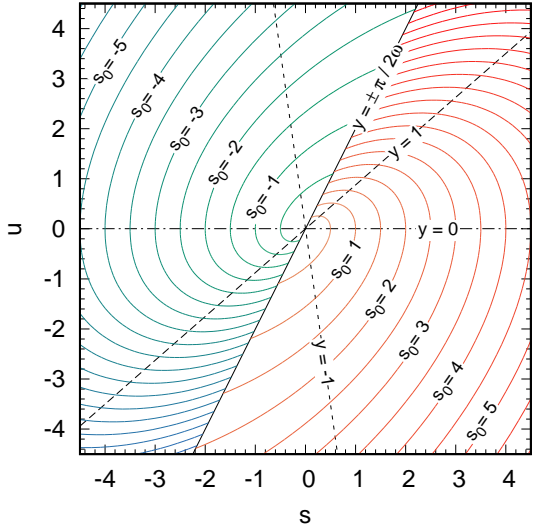}
    \caption{Characteristic curves in the $(s,u)$ plane for Eq.~\eqref{eq:statchar_app}. Different colors correspond to different values of $s_0$. Black dashed lines identify some $y=const$ curves; the continuous black line $y=\pm \pi/2 \omega$ separates the half-planes with $s_0<0$ and $s_0>0$. Here $\omega=1$.}
    \label{fig:applot}
\end{figure}

First, we have to find explicit expressions for $u(y)$ and $s(y)$ along the infinite characteristic curves determined by Eqs.~\eqref{eq:odesu}.
From those relations, one derives the second order differential equation
\begin{equation}
\label{eq:secondode}
    \frac{d^2u}{dy^2} + \frac{du}{dy}+\stud u=0\,,
\end{equation}
which is solved by
\begin{equation}
\label{eq:uy}
    \widetilde{u}(y;s_0)=\frac{s_0}{\omega}\sin(\omega y) e^{-y /2}=\frac{s_0}{\omega}g(y) \,,
\end{equation}
where $s_0$ is a parameter whose value discriminates between different curves, and we have introduced the angular frequency of the damped oscillator,
\begin{equation}
\omega=   \sqrt{\stud-\frac{1}{4}}\,.
\end{equation}
We will assume that $\omega$ is real, since we are interested in the underdamped limit. We have also introduced the function
\begin{equation}
    g(y)=\sin(\omega y) e^{-y /2}\,.
\end{equation}
Of course, Eq.~\eqref{eq:secondode} is also solved by any function of the kind
\begin{equation}
  \widetilde{u}(y;s_0,y_0)=\frac{s_0}{\omega}g(y-y_0)\,,  
\end{equation}
obtained by shifting the argument of the solution~\eqref{eq:uy} by an arbitrary constant $y_0$. However, all of them describe the same characteristic curve in the $(s,u)$ plane, up to an irrelevant change of parametrization, so that we can safely impose $y_0=0$.
The second of Eqs.~\eqref{eq:odesu} implies
\begin{equation}
\label{eq:sy}
    \widetilde{s}(y;s_0)=\cbr{\frac{1}{2}+\frac{\omega}{\tan(\omega y)}}u(y)\,.
\end{equation}
The curves identified by $(s(y;s_0),u(y;s_0))$, for given values of $s_0$, are represented in Fig.~\ref{fig:applot}. When $y=0$, each curve crosses the $s$ axis, and $s=s_0$. For $y \to \pm \pi/2\omega$ the curve approaches the $u=2m/ s$ line.

We can now solve Eq.~\eqref{eq:odechar}, which is a linear homogeneous ordinary differential equation with non-constant coefficients. The solution is expressed as
\begin{equation}
\label{eq:solpy}
\begin{aligned}
    \widehat{p}_{st}(y;s_0)&=F(s_0)\exp \int_0^y dy' \cbr{-i\stud \mean \widetilde{u}(y';s_0)+|\widetilde{u}(y';s_0)|^{\alpha}}\\
    &= F(s_0)\exp \int_0^y dy'\cbr{ -i\stud \mean \frac{s_0 g(y')}{\omega}+\Big|\frac{s_0g(y')}{\omega}\Big|^{\alpha}}
    \,,
\end{aligned}
\end{equation}
where $F(s_0)$ is an arbitrary function of $s_0$, and we have made use of Eq.~\eqref{eq:uy}. At this point we only have to substitute the pair $(s_0,y)$ with the corresponding $(s,u)$, by inverting Eqs.~\eqref{eq:uy} and~\eqref{eq:sy}. It is found that
\begin{subequations}
\begin{equation}
\label{eq:defy_app}
    y(s,u)=\frac{1}{\omega}\arctan\cbr{\frac{\omega}{\frac{s}{u}-\frac{1}{2}}}
\end{equation}
\begin{equation}
    s_0(s,u)=\frac{\omega u}{g(y(s,u))}\,.
\end{equation}
\end{subequations}
Equation~\eqref{eq:solpy} can be rewritten as
\begin{equation}
\label{eq:steppstat}
 \widehat{p}_{st}(s,u)=F\cbr{s_0}\exp\sbr{-i\stud \mean u\mathcal{G}_1(y)   +  |u|^{\alpha}\mathcal{G}_{\alpha}(y)}
\end{equation} 
where $y=y(s,u)$, $s_0=s_0(s,u)$ and
\begin{equation}
    \mathcal{G}_{a}(y)=\frac{1}{[g(y)]^a}\int_0^y dy'[g(y')]^a\,.
\end{equation}

We still have to impose the functional form of $F$. The normalization condition $\widehat{p}_{st}(0,0)=1$ only implies $F(0)=1$. In order to have enough constraints, we should also require $p(x,v)$ to be always positive, and vanishing for $x,v \to \pm \infty$. This condition is quite difficult to implement; instead, one may impose that the marginalized stationary distribution for the particle positions is the same as in the overdamped limit. This marginal distribution can be written as
\begin{equation}
   p_{st}(x)=\frac{1}{2 \pi} \int_{-\infty}^{\infty}ds e^{isx}\widehat{p}(s,0)=\frac{1}{2 \pi} \int_{-\infty}^{\infty}ds e^{isx}F(s)\,,
\end{equation}
where we have used the fact that $s_0(s,u)\to s$ when $u\to0$.
One obtains
\begin{equation}
    \widehat{p}_{st}(s,0)=\widehat{p}_{st}(s)=\exp\cbr{i\mean s-\frac{|s|^{\alpha}}{{\alpha}\stud}}\,;
\end{equation}
as a consequence, it can be concluded by comparison that
\begin{equation}
\label{eq:stepnorm}
    F(s_0)=\exp\cbr{i\mean s_0-\frac{|s_0|^{\alpha}}{\alpha  \stud }}\,.
\end{equation}

Finally, let us notice that
\begin{equation}
\begin{aligned}
    \mathcal{G}_1(y)&=\frac{e^{y /2}}{\sin(\omega y)} \int_0^{y}dy'\sin(\omega y')e^{- y' /2}\\
    &=\frac{1}{\omega^2+1/4}\cbr{\frac{\omega e^{y /2}}{\sin(\omega y)} - \frac{\omega}{\tan(\omega y)} - \frac{1}{2}  }\\
    &=\frac{\omega}{\stud g(y)}- \frac{s}{\stud u}\,,
    \end{aligned}
\end{equation}
where in the last step we have made use of Eq.~\eqref{eq:defy_app}. Inserting this result into Eq.~\eqref{eq:steppstat}, and taking into account Eq.~\eqref{eq:stepnorm}, a simpler expression for the characteristic function of the stationary distribution can be obtained:
\begin{equation}
     \widehat{p}_{st}(s,u)=\exp\sbr{i\mean s   +  |u|^{\alpha}\mathcal{G}_{\alpha}(y)-\frac{|s_0|^{\alpha}}{\alpha k}}\,,
\end{equation}
where all terms depending on $s_0$ have been absorbed into $F(s_0)$. The functional form of $F(s_0)$ may be fixed by passing to real space and imposing proper boundary conditions for the PDF. However, as discussed in the main text, this is not needed for our purposes.

\section{Properties of the operator $L$}
\label{sec:appprop}
In this appendix, we show two properties of the operator $L$ defined by Eq.~\eqref{eq:linearopl}, namely Eq.~\eqref{eq:propl1} and~\eqref{eq:propl2}.

First, let us compute two quantities whose explicit expression will be useful for the following derivation:
\begin{equation}
    g'(y) =\cbr{\frac{\omega}{\tan(\omega y)}-\frac{1}{2}}g(y)=\cbr{\frac{s}{u}-1}g(y)
\end{equation}
and
\begin{equation}
\label{eq:propdsy}
    \partial_s y =\frac{1}{ s- \frac{s^2}{u}-\frac{u}{4} - \omega^2 u}=\cbr{s-\frac{s^2}{u} - \stud u}^{-1}\,.
\end{equation}
Let us also notice that Eq.~\eqref{eq:defy_app} implies 
\begin{equation}
u\partial_u y = - s \partial_s y\,.
\end{equation}

Recalling definition~\eqref{eq:defs0} and taking into account the above results, it is immediate to show that, for a generic function $f(s_0)$,
\begin{equation}
\begin{aligned}
    L[f(s_0)]&=\sbr{s-u-\frac{ s u-s^2 - \stud u^2}{g(y)} g'(y)\, \partial_s y}\frac{\omega f'(s_0)}{g(y)}\\
    &=\sbr{s-u-u\cbr{\frac{s}{u}-1} }\frac{\omega f'(s_0)}{g(y)}=0\,,
\end{aligned}
\end{equation}
which is nothing but Eq.~\eqref{eq:propl1}.

Finally, let us compute
\begin{equation}
    \begin{aligned}
    L[u^a\mathcal{G}_a(y)]=&\cbr{s-u} au^{a-1}\mathcal{G}_a(y) +\\&+u^a\sbr{\cbr{s-\frac{s^2}{u}-\stud u}\partial_s y}  \mathcal{G}_a'(y)\,.
    \end{aligned}
\end{equation}
The term in square parentheses is equal to 1, due to Eq.~\eqref{eq:propdsy}. By noticing that
\begin{equation}
\begin{aligned}
    \mathcal{G}_a'(y)&= -\frac{a g'(y)}{[g(y)]^{a+1}}\int_0^y dy' g^a(y')  + 1\\
    &=-a\cbr{\frac{s}{u}-1}\mathcal{G}_a(y) + 1\,,
    \end{aligned}
\end{equation}
one gets
\begin{equation}
   L[u^a\mathcal{G}_a(y)]=u^a\,, 
\end{equation}
i.e. Eq.~\eqref{eq:propl2}.

\section{The underdamped Brownian case}
\label{sec:appundbrown}
This appendix is devoted to the study of the Brownian case $\alpha=2$. In this case the proposed ansatz has an explicit expression also in real space, and it can be checked that it corresponds to the known solution of the Fokker-Planck equation for the dynamics.

Our ansatz~\eqref{eq:udansatz}, taking into account the condition~\eqref{eq:condundh}, reads in the Brownian case
\begin{equation}
\label{eq:logansatz2}
    \ln \widehat{p} = i \tmean s + i \dot{\tmean} u + u^2\mathcal{G}_2(y) -\frac{s_0^{2}}{2 k}\,.
\end{equation}
Let us compute $\mathcal{G}_2$ explicitly:
\begin{equation}
    \begin{aligned}
    \mathcal{G}_2(y)&=\frac{e^{ y/m}}{\sin^2(\omega y)}\int_0^y\,dy' e^{- y'/m}\sin^2(\omega y')\\
    &=\frac{m^2\omega^2\cbr{e^{ y/m}-1}}{2  \stud \sin^2(\omega y)}-\frac{m\omega}{2 \stud \tan(\omega y)} - \frac{1}{4 \stud}\,,
    \end{aligned}
\end{equation}
where we have made use of the identity $1+4\omega^2m^2=4m\stud$.
Once inserted into Eq.~\eqref{eq:logansatz2}, the above relation leads to
\begin{equation}
    \begin{aligned}
    \ln \widehat{p} =& i \tmean s + i \dot{\tmean} u-\frac{\omega^2 u^2}{2  \stud \sin^2(\omega y)}-\frac{\omega u^2 }{2 \stud \tan(\omega y)}-\frac{ u^2}{4 \stud}\\
    = &i \tmean s + i \dot{\tmean} u-\frac{ u^2}{2 }-\frac{ s^2}{2 \stud} \,.
    \end{aligned}
\end{equation}
In the last step we have exploited the definition of $y$, Eq.~\eqref{eq:defy_app}.

At this point it is possible to write explicitly the probability density function of the particle in real space. Indeed
\begin{equation}
    \begin{aligned}
    p(x,v,t)&=\frac{1}{4 \pi}\int_{-\infty}^{\infty}ds\,e^{-is(x-\tmean)-\frac{ s^2}{2 \stud}} \int_{-\infty}^{\infty}du\,e^{-iu(v-\dot{\tmean})- \frac{u^2}{2}}\\
    &=\frac{\sqrt{\stud}}{2 \pi }e^{-\frac{(v-\dot{\tmean})^2}{2}-\frac{ \stud}{2 }(x-\tmean)^2}\,.
    \end{aligned}
\end{equation}
Let us notice that this solution is consistent with the expected shape for the (equilibrium) stationary state, given in this case by a Maxwell-Boltzmann distribution when $\dot{\tmean}=0$. We have now to check that the above ansatz, once plugged in the Fokker-Planck equation
\begin{equation}
    \partial_t p = - \partial_x (v p) + \partial_v \sbr{\stud (x-\mean) p +  v p} + \partial_v^2 p
\end{equation}
leads to the correct condition. Indeed one obtains
\begin{equation}
    (v  -\dot{\tmean})\cbr{  \stud \tmean + \dot{\tmean} + \ddot{\tmean}  - \stud \mean  }p=0\,,
\end{equation}
which implies Eq.~\eqref{eq:condund}, as expected.

\bibliography{biblio}

\end{document}